# Gravitational and Cosmological Spectral Shift with Remote Quantum States

Charles Francis

*25 Elphinstone Rd., Hastings, E. Sussex, TN34 2EG*

A class of coordinate systems is found for Friedmann Cosmologies with local gravity such that it is possible to formulate quantum theory over astronomical and cosmological distances. When light from distance objects is treated as a quantum motion, new predictions are found for cosmological redshift and lensing. Good agreement is found between predictions and supernova redshifts for a closed Friedmann Cosmology with no cosmological constant and expanding at half the rate of the standard model. A previously unmodelled component of cosmological redshift accounts for the anomalous Pioneer blueshift, and for the flattening of galaxy rotation curves simulating a MONDian law. Distant lenses have a quarter of the mass required by standard general relativity. Missing mass can be accounted by a massive neutrino. CDM is not required.

## 1 Introduction

A connection describes the parallel displacement of vector quantities between nearby points in a classical manifold. This paper identifies a particular class of coordinate systems, *quantum coordinates*, in which it is possible to formulate quantum theory with remote initial and final states in Friedmann cosmologies with local gravity. The parallel displacement in quantum coordinates of momentum from the initial state to the remote final state will be called a *teleconnection*. The teleconnection reduces to the Levi-Civita connection for small displacements between initial and final states. Standard general relativity and quantum mechanics are assumed, but light from distant astronomical objects is treated quantum mechanically, as photons. This will lead to changes in predictions concerning cosmological redshift and bending of light, but to no other changes to the predictions of general relativity.

The treatment assumes an interpretation of quantum theory in which the wave function is a means of calculating probabilities for the results of measurement. The quantum state describes knowledge of the particle. "Knowledge" in this context refers to information which is available in principle from the physical situation, whether or not it is known in practice by a particular observer. The knowledge about a particle which is possible in principle depends on the physical relationship of that particle with other matter, and, in the context of an expanding universe, will have empirical consequences which cannot be reconciled with a view of the wave function as a physical field. In particular, the direct relationship between frequency and energy will be lost when dealing with wave functions from a distant source.

Section 2 describes the teleconnection in a Friedmann cosmology. A square law is shown for cosmological redshift (section 2.2), but energy transfer is as in classical general relativity (section 2.3). The anomalous Pioneer blueshift is found in section 2.5. Section 3 describes cosmological implications. Under the square law, the rate of expansion is a half, and critical mass for closure is a quarter, of that of the standard model (section 3.1). The luminosity redshift relation is calculated (section 3.2). Supernova redshifts are consistent with a closed Friedmann cosmology with roughly twice critical mass and no cosmological constant (section 3.3). The improvement in fit compared to the concordance model is not significant. The square redshift law resolves possible issues with galaxy ageing and makes a clear prediction that new telescopes, such as Herschel, will find many large high-redshift galaxies (section 3.4). There is no major predicted change in cosmic microwave background (CBR) or big bang nucleosynthesis (BBN), but revisions to peculiar motions in the galactic neighbourhood will have an impact on CBR anisotropy (section 3.5).

Section 4 extends the treatment to Friedmann Cosmologies with local fluctuations in gravity. The form of the metric for static coordinates in found in section 4.1. Expansion is incorporated in section 4.2. Gravitational redshift is not affected by the quantum treatment (section 4.3). The Levi-Civita connection is found in the classical correspondence (section 4.4). Distant gravitational lenses have a quarter of the mass required in standard cosmology, although local bending of light around the Sun is treated classically and is not affected. (section 4.5). Quantum coordinates are defined in section 4.6. Galactic rotation curves are Newtonian, but display an apparent flattening which replicates a MONDian law (section 4.7). Conclusions are summarized in section 5.

## 2 Friedmann Cosmologies

### 2.1 The Teleconnection in a Friedmann Cosmology

Using $\tau$-$\rho$ coordinates for a Penrose diagram (Penrose coordinates), in a Friedmann cosmology the metric is

$$ds^2 = a^2(\tau)(d\tau^2 - d\rho^2 - f^2(\rho)(d\theta^2 + \sin^2\theta d\phi^2)),$$

where $f(\rho) \in \{\sin\rho, \rho, \sinh\rho\}$ for space with positive, zero or negative curvature respectively. Coordinate time, $\tau$, is related to cosmic time, $t$, by $a(\tau)d\tau = dt$. An observer, Alf, at A at cosmic time $t_1$ and coordinate time $\tau_1$, defines radial locally Minkowski coordinates, $(t, r, \theta, \phi)$ based on cosmic time, $t$, such that $dt = a(\tau)d\tau$ and $r = a(\tau)\rho$. The metric is

$$ds^2 = dt^2 - dr^2 - a^2(t)f^2\left(\frac{r}{a(t)}\right)(d\theta^2 + \sin^2\theta d\phi^2).$$

For $r \ll a$,

$$ds^2 \approx dt^2 - dr^2 - r^2(d\theta^2 + \sin^2\theta d\phi^2).$$

Alf defines a non-physical metric, $\bar{h}(x)$, using τ-ρ coordinates, such that,

$$d\bar{\sigma}^2 = A^2(d\tau^2 - d\rho^2) - B^2 f^2(r)(d\theta^2 + \sin^2\theta d\phi^2).$$

where $A$ and $B$ are real numbers, whose values are to be determined. Alf defines barred vectors in τ-ρ coordinates with non-physical metric $\bar{h}$.

**Definition:** *For a vector $x = (x^\tau, x^\rho, x^\theta, x^\phi)$ at $(\tau_1, A)$, the corresponding barred vector is*

$$\bar{x} = \left(\frac{x^\tau}{A}, \frac{x^\rho}{A}, \frac{x^\theta}{B}, \frac{x^\phi}{B}\right).$$

This definition will apply to classical energy-momentum, but not to quantum energy-momentum which appears in the wave function. Barred quantum energy-momentum will be defined separately.

**Definition:** *For the displacement, $x = (x^\tau, x^\rho, x^\theta, x^\phi)$, from $(\tau_1, A)$, the corresponding barred displacement vector is*

$$\bar{x} = \left(\frac{x^\tau}{A}, \frac{x^\rho}{A}, \frac{x^\theta}{B}, \frac{x^\phi}{B}\right).$$

It follows that for a vector (including a displacement vector) $x = (x^t, x^r, x^\theta, x^\phi)$, in locally Minkowski coordinates at $(t_1, A)$,

$$\bar{x} = \left(\frac{x^t}{aA}, \frac{x^r}{aA}, \frac{x^\theta}{B}, \frac{x^\phi}{B}\right),$$

and that, for barred vectors $\bar{x}$ and $\bar{y}$ at $(\tau_1, A)$, the barred dot product, evaluated with non-physical metric $\bar{h}$, satisfies

$$a^2(\tau_1)\bar{x} \cdot \bar{y} = x \cdot y.$$

Alf formulates quantum states locally in Hilbert space at time $t_1$, and defines plane wave states at $t_1$ using

$$\langle x|p\rangle = \left(\frac{1}{2\pi}\right)^{3/2} e^{ix \cdot p}.$$

Quantum theory is then reformulated in terms of barred quantities, under the requirement that the inner product is preserved.

$$\langle \bar{x}|\bar{p}\rangle = \left(\frac{1}{2\pi}\right)^{3/2} e^{-i\bar{x} \cdot \bar{p}} = \left(\frac{1}{2\pi}\right)^{3/2} e^{-ix \cdot p} = \langle x|p\rangle.$$

This requires that $\bar{x} \cdot \bar{p} = x \cdot p$, and we define momentum in quantum theory:

**Definition:** *Barred momentum in quantum theory is*

$$\bar{p} = \left(\frac{ap^t}{A}, \frac{ap^r}{A}, \frac{p^\theta}{B}, \frac{p^\phi}{B}\right).$$

It follows that the momentum space wave function,

$$\langle \bar{p}|f \rangle = \left(\frac{1}{2\pi}\right)^{3/2} \int_{\bar{r}=0}^{\bar{r}=\pi/A} d^3\bar{x} \langle \bar{x}|f \rangle \, e^{i\bar{x}\cdot\bar{p}}$$

$$= \left(\frac{1}{2\pi}\right)^{3/2} \int_{r=0}^{r=\pi a} d^3x \langle x|f \rangle \, e^{ix\cdot p}$$

$$= \langle p|f \rangle,$$

is preserved when it is given that $\langle x|f \rangle$ vanishes outside a sphere of radius $\pi a$, and when the wave function is normalised such that $\langle \bar{x}|f \rangle = a^3 A^3 \langle x|f \rangle$.

A connection defines the notion of parallel in vector spaces defined at nearby points of a manifold. The teleconnection defines the parallel displacement of barred momentum in quantum mechanics from an initial state to a final state when the reference matter used to describe the initial state is remote from that used to describe the final state. It will be shown that when the initial and final states are determined with respect to nearby reference matter, the teleconnection is equivalent to the Levi-Civita connection. The teleconnection defines the inner product between the initial state of a particle emitted at $(t_1, A)$ and the final state of that particle detected some later time.

**Definition:** *For the barred momentum $\bar{p}$, the plane wave state in $\tau$-$\rho$ coordinates with non-physical metric $\bar{h}$ at any time is*

$$\langle \bar{x}|\bar{p} \rangle = \left(\frac{1}{2\pi}\right)^{3/2} e^{-i\bar{x}\cdot\bar{p}},$$

*where the barred dot product uses the non-physical metric, $\bar{h}$.*

This definition preserves Newton's first law and the constancy of the speed of light in $\tau$-$\rho$ coordinates with non-physical metric $\bar{h}$. It is required to retain the formal structure of relativistic quantum theory, and preserves the momentum space wave function as a constant of the motion. The evolution of the wave function is determined in $\tau$-$\rho$ coordinates with non-physical metric $\bar{h}$, and is such that it corresponds to the standard wave function in locally Minkowski coordinates.

### 2.2 Cosmological Redshift

A second observer, Beth, at $(\tau_0, B)$, remote from $(\tau_1, A)$ where $\tau_0 > \tau_1$, also defines quantum states in Hilbert space.

**Theorem:** Let $a_1 = a(t_1)$ and $a_0 = a(t_0)$. Light emitted at time $t_1$ with wavelength $\lambda_1$ from a distant point A, and detected at B, at time $t_0$ with wavelength $\lambda_0$ is redshifted according to

$$\lambda_0 = (1+z)\lambda_1 = \frac{a_0^2}{a_1^2}\lambda_1,$$

**Proof:** One period of light in locally Minkowski coordinates, $(t, r, \theta, \phi)$, with an origin at A at cosmic time $t_1$ is represented by a timelike vector of magnitude $\lambda_1$ (figure 1). In $\tau$-$\rho$ coordinates, its timelike component is $\lambda_1/a_1$. The corresponding barred quantity is $\lambda_1/Aa_1$. The barred vector is translated to $(\tau, B)$. The corresponding unbarred quantity is found from the physical metric, $g$, and has magnitude $a_0\lambda_1/a_1^2$. On transforming to primed locally Minkowski coordinates $(t', r', \theta', \phi')$, with an origin at B, we find

$$\lambda_0 = \frac{a_0^2}{a_1^2}\lambda_1.$$

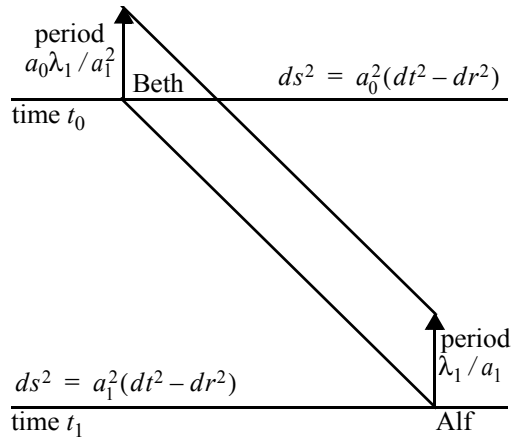

**Figure 1:** Parallel Displacement in $\tau$-$\rho$ coordinates

It follows that, for small $r$,

$$1 + z \approx 1 + 2r\dot{a}/a.$$

Thus recession velocity due to expansion is half the value calculated from Doppler, and coordinates in which radial distance from Earth is calculated from redshift exhibit a stretch of factor half in the radial direction. So $A = 2$. The time taken for a pulse of light at this distance to traverse a small angular distance $d\theta$ is $rd\theta = \bar{r}d\bar{\theta} = rd\theta/AB$. So $B = 1/2$ (a stretch of factor two in angular directions gives $4\pi$ in a circle, which may be related to the spin states of Fermions). Thus, the non-physical metric, $\bar{h}$, is

$$d\bar{\sigma}^2 = 4(d\tau^2 - d\rho^2) - \frac{1}{4}f^2(\rho)(d\theta^2 + \sin^2\theta d\phi^2). \tag{2.2.1}$$

### 2.3 Energy Transfer

The squared redshift law appears at first to be at odds with the claim that parallel displacement under the teleconnection reduces to parallel transport in the classical correspondence. This is due to the manner in which expansion is treated in the quantum theory and is resolved at the point of the collapse of the wave function. The square redshift law applies to spectroscopic measurements of light from distant sources, when the detection of the photon takes place *after* diffraction or refraction of the quantum wave function. It does not affect energy transfer because, in order to formulate the inner product between an initial state at time $t_0$ and a final state at time $t_1$, Beth enlarges the coordinate axes at time $t_0$ by a factor $a_0/a_1$. The square law applies to spectroscopic measurements, but, since the initial measurement of energy-momentum is relative to the coordinate axes at time $t_0$, this factor must be removed from the calculation of energy-momentum of light from a distant source. Thus, the energy measured locally by Beth at time $t_0$ of a photon emitted at time $t_1$ with energy from a from a distant source at time $t_1$ is given by

$$p_0^0 = \frac{A}{a_0}\bar{p}^0 = \frac{a_1}{a_0}p_1^0,$$

as in classical general relativity.

### 2.4 Suppression of Expansion in a Neighbourhood

The inner product is defined to generate probabilities. To specify a probability, it is necessary to specify the known conditions to which the probability applies. We may interpret the collapse of the wave function as the change in probability when the known condition changes. The teleconnection will be applied to two types of physical situation, depending on whether the clock used to determine the final state can be calibrated to the clock used to determine the initial state, for example by Einstein's calibration procedure.

1. When there is a physical calibration between the clocks used to measure initial and final states quantum theory is formulated without expansion.
2. When no calibration is possible, as is the case for light from a distant stellar object, the quantum theory must be formulated taking into account expansion between coordinates used for initial and final states.

Thus, the teleconnection gives two distinct laws for redshift depending on whether processes used for the description of the initial state can be calibrated to those used for the final state. It is not clear precisely how to state the conditions under which each law applies, but the dichotomy will be seen both in the onset of the Pioneer anomaly (section 2.5) and in the analysis of galaxy rotation curves (section 4.7). It should be not be understood as a change in physical law, but as a change in the mathematical formulation of quantum theory as a theory of probabilities, reflecting the difference in information available to the observer depending on whether synchronisation is possible between the clocks used to determine the initial and final states.

Because a factor of the expansion parameter is removed in the calculation of energy/momentum (section 2.3), both formulations will give the same classical quantities, but there will be a difference in predictions for spectral shifts, determined, for example, by diffraction, prior to the actual detection of a particle from a distant astronomical source. This is legitimate for interpretations of quantum theory in which the wave function is a device for the calculation of probabilities, but appears to exclude interpretations which attribute any form of ontological existence to the wave function.

At the present time, it is not possible to write down conditions under which Einstein's synchronisation procedure is possible. It is known that the anomalous Pioneer blue shift appeared after radar lock was lost. It is not known why radar lock was lost. This is usually put down to equipment failure due to unknown cause, possibly hostile conditions during the Jupiter flyby. After radar lock was lost, it was no longer possible to calibrate processes on Pioneer to processes on Earth. Synchronisation became impossible in practice, creating the conditions under which the anomalous shift was observed. It remains to be established whether the anomaly appears in consequence of some deep reason dependent on local geometry and motion according to which maintenance of radar lock is impossible in principle, or whether it is sufficient that a synchronisation procedure cannot be carried out in practice. A quantum theoretic explanation would be more satisfying if it were possible to present some deep underlying reason of principle why radar should be lost at a given point, rather than simple equipment failure. No such reason of principle has been found, but if there is such a reason it might be expected to involve the distance of a radiation source from the Sun as well as the peculiar motions of both the Sun and the radiation source relative to $\tau$-$\rho$ coordinates. Hopefully future missions will cast greater light on the onset of the anomaly.

## 2.5 Anomalous Pioneer Blueshift

In the absence of calibration between clocks on a distant body and clocks on the earth, the momentum of the distant body is constant in τ-ρ coordinates with non-physical metric $\bar{h}(x)$. Ignoring local gravitational effects (set $k = 1$ in the metric, 4.3.1), as determined in signals detected on Earth, classical energy is given by (note that $t_0$ is now the earlier time)

$$p^0(t) = (1+z)p^0(t_0) = \frac{a^2}{a_0^2}p^0(t_0) \approx 1 + (t-t_0)H_0 p^0(t_0).$$

Classical energy is proportional to the rate of clocks on the distant body, so signals show a frequency drift, $H_0$, toward the blue.

This effect appears to have been observed in the anomalous Pioneer blueshift. For some years the Pioneer spacecraft sent back Doppler information interpreted as an anomalous acceleration toward the sun (Anderson et al., 2002). Although what is measured is blueshift, JPL expressed their result in the form of an equivalent classical acceleration, $a_P = 8.74 \pm 1.33 \times 10^{-8} \text{cms}^{-2}$. Despite thorough analysis of conceivable possibilities by JPL, no explanation has been found in classical physics for the anomalous blueshift.

The blueshift can also be expressed as a drift in frequency, $2.92 \pm 0.44 \times 10^{-18}$ s/s$^2$. JPL have commented that this value is close to Hubble's constant, suggesting a possible cosmological origin for the shift. In the standard units for Hubble's constant the drift is $90 \pm 14$ km s$^{-1}$ Mpc$^{-1}$. The drift was observed only after radar lock was lost with Pioneer, in accordance with the prediction that it is only present while there is no means of synchronisation of clocks on Pioneer with those on Earth. A preliminary analysis of planetary flybys also appears to show an anomaly during a period when radar lock is lost (Anderson et al, 2006). If the explanation given here is correct, then there is no corresponding classical acceleration and, if its position could be measured by ranging, Pioneer would be found on the expected Newtonian path.

## 3 Cosmological Implications

### 3.1 Cosmological Parameters

The teleconnection replaces the standard linear redshift law with a square law. So,
$$1 + z = a^2(t)/a^2(t-r) \approx a^2(a - r\dot{a})^{-2} \approx 1 + 2r\dot{a}/a,$$
from which we read the value of Hubble's parameter,
$$H = 2\dot{a}/a = 2\dot{a}(t)/a(t).$$
This differs by a factor of 2 from standard general relativity. It follows immediately that the rate of expansion of the universe is half, and that, for given cosmological parameters, the universe is twice as old as would be indicated by the standard redshift law.

Friedmann's equation is
$$\frac{\dot{a}^2}{a^2} = \frac{8\pi G\rho}{3} - \frac{k}{a^2} + \frac{\Lambda a^2}{3}.$$

Normalising, such that $\Omega = 1$ is critical density for closure in a no-$\Lambda$ model, $\Omega$ takes on four times its standard value, and critical density for closure is a quarter of that of standard general relativity.

$$\Omega = \frac{32\pi G\rho_0}{3H_0^2}, \quad \Omega_k = -\frac{4k}{H_0^2 a_0^2}, \quad \Omega_\Lambda = \frac{4\Lambda}{3H_0^2}, \tag{3.1.1}$$

where $k = -1, 0, 1$ for a space of negative, zero or positive curvature respectively. With these definitions, Friedmann's equation is

$$\dot{a}/a = \tfrac{1}{2}H_0(\Omega(1+z)^{3/2} + \Omega_k(1+z) + \Omega_\Lambda)^{1/2},$$

requiring that $\Omega + \Omega_k + \Omega_\Lambda = 1$ as usual.

### 3.2 The Luminosity-Redshift Relation

The calculation of luminosity distance under the teleconnection follows the same pattern as in the standard model, but using the squared redshift law. The coordinate distance from emission at time $t_e$ to detection at time $t_0$ is

$$\rho = \int_{t_e}^{t_0} \frac{1}{a}dt = \int_{a_e^2}^{a_0^2} \frac{1}{2a^2\dot{a}}d(a^2) = \int_0^z \frac{k\Omega_k^{1/2}(1+z)^{3/2}}{2(\Omega(1+z)^{3/2} + \Omega_k(1+z) + \Omega_\Lambda)^{1/2}}dz.$$

Angular size distance is $a_0 \sin 2\rho$ taking into account the doubling of angles required by the non-physical metric. The determination of luminosity distance now proceeds as in the standard model, observing that energy and rate of photon emission are shifted by $1+z$, as is standard. Thus, luminosity distance is

$$d_L = \frac{c(1+z)}{H_0\Omega_k^{1/2}}\sin 2\rho = \frac{c(1+z)}{H_0\Omega_k^{1/2}}\sin\int_0^z \frac{k\Omega_k^{1/2}(1+z)^{3/2}}{(\Omega(1+z)^{3/2} + \Omega_k(1+z) + \Omega_\Lambda)^{1/2}}dz.$$

The distance modulus is $\mu = 5\log d_L + 25$ as usual.

### 3.3 Supernova Redshift

All Type 1A supernovae are generated from a near identical process, and have almost exactly the same intrinsic brightness, or absolute magnitude. Redshift and visual magnitude each give a measure of the distance of the supernova. The Union compilation, prepared by the Supernova Cosmology Project (Kowalski et al., 2008) contains, at the time of writing, the most up to date supernova data from different sources, prepared in as uniform a manner as possible, from which 307 Type 1A supernovae pass usability tests. A best fit to the data is found by adjusting cosmological parameters so as to minimize $\chi^2$, the sum of normalised squared differences between the data points and the curve.

Data on the cosmic microwave background determined from the Wilkinson Microwave Anisotropy Probe indicates a flat space ($\Omega_k = 0$) model with non-zero cosmological constant, known as the concordance model. For 307 SNe in the Union compilation, the best fit concordance model gives $\Omega = 0.29$ and $\chi^2 = 311.0$. The best fit teleconnection no-$\Lambda$ model gives $\Omega = 1.92$ and $\chi^2 = 309.4$. The difference in $\chi^2$ values is too small for comparison of the quality of the fits, as can also be seen in the plots (figure 2 and figure 3). The predicted curves for the teleconnection no-Λ model and the standard concordance model are sufficiently close that it would only be possible to distinguish them by taking a large sample of supernovae with redshifts greater than about 1.4. The Joint Dark Energy Mission is currently considering three concepts for space telescopes intended to collect the sort of data which would be required, with projected launch dates before 2020.

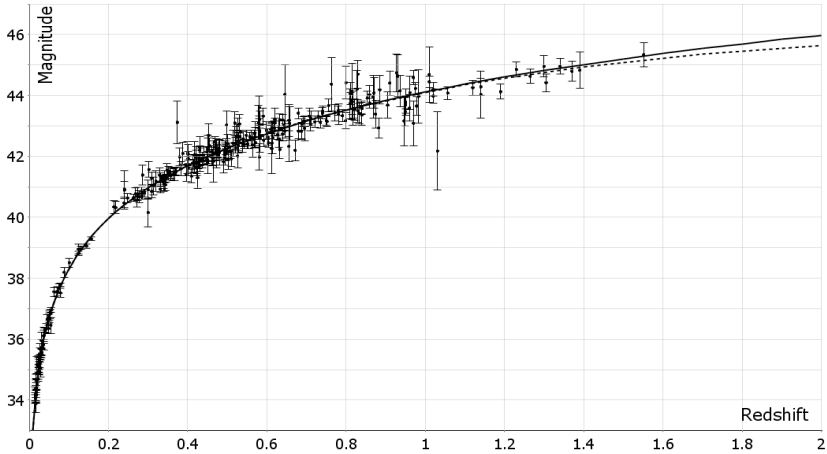

**Figure 2:** Magnitude against redshift for 307 supernovae from the Union compilation. Theoretical curves for the best fit concordance model (black) and teleconnection models (dashed) are also plotted.

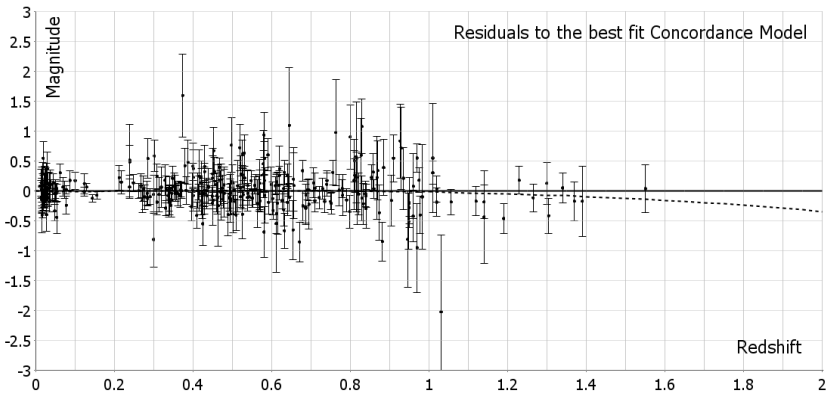

**Figure 3:** Residuals to the best fit concordance model.

### 3.4 Galaxy Ageing

If observations at high redshift had revealed the expected activity of the early universe it would have falsified the square redshift law; in fact it receives support from the observation of mature galaxies at $z = 1.4$ and greater (e.g. Mullis et al., 2005; Doherty et al., 2005, and references cited therein). As described by Glazebrook (2004), there is poor agreement between current theoretical models of galaxy evolution and empirical data. No-CDM models predict large galaxies form in the early universe, while CDM predicts hierarchical galaxy formation, but there does not appear to be sufficient time for galaxies to form at observed redshifts (especially for no-CDM models). To explain this it has been suggested (e.g. Cimatti et al., 2004) that the theoretical models may be inaccurate. The problem is exacerbated by recent calculations based on Beryllium abundancies by Pasquini et al. (2004) which give an age of $13.4 \pm 1$ Gyr for a globular cluster formed about 0.2 Gyr after the Milky Way. The teleconnection presents an alternative, that a square redshift law means we have to revise the ages of red galaxies. A value of Hubble's constant $h = 0.71$ together with $\Omega = 1.9$ gives the age of the universe as 15.7 Gyr and at redshift 6 the universe would have been about 3.2 Gyr. A detailed study is required to assess consistency between observation and theory, but if the square law of cosmological redshift is correct, then observations with the recently launched Herschel Space Observatory will show many large galaxies at redshifts considerably too high to be explained in standard cosmology.

### 3.5 Big Bang Nucleosynthesis and CBR

The cosmological microwave background is continuously observable, so that the standard linear redshift law applies to the CBR. Since the age of the universe is close to that of the standard model, the initial rate of expansion is also similar and the analyses of big bang nucleosynthesis and of decoupling are unaltered. Based on figures from Liddle (2003), the density of baryonic matter becomes $0.064 \leq \Omega_B h^2 \leq 0.096$ after normalising $\Omega_{cr}$ to 1 according to (3.1.1). Thus baryonic matter forms 10-20% of critical density. For $\Omega \approx 2$ the ratio of non-baryonic to baryonic matter need only be 9:1, and, after taking into account the redshift-age relation, the prospect remains open that this might be accounted for by a massive neutrino.

The concordance model is supported by the integrated Sachs-Wolfe effect (Afshordi, Loh & Strauss; 2004; Boughn & Crittendon, 2004; Fosalba et al., 2003; Nolta et al., 2004; Scranton et al., 2004) using evidence from the Two-Degree Field Galaxy Redshift Survey (2dFGRS; Peacock et al., 2001; Percival et al., 2001; Efstathiou, 2002), and from the Wilkinson Microwave Anisotropy Probe (WMAP; Spergal, 2003, and references cited therein). In practice these measurements determine cosmological parameters rather than test consistency, and they depend on the distance-redshift relation. Acceleration depends only on distance and time, so that, if the standard model is consistent, a change in the distance-redshift relation can be expected to give a consistent change in the deceleration parameter in different tests in which distance is found from redshift. Thus it is to be expected that $\Omega_s \approx 0.3$ corresponds to $\Omega \approx 2$ in the teleconnection model whether it is determined from Supernova or from WMAP and 2dFGRS.

The broad properties of the microwave background are unchanged in the teleconnection model, as we expect isotropy and a Gaussian random distribution, but it is not possible without detailed analysis to interpret the WMAP data as implying that $\Omega_k = 0$, since this may be modified in analysis. McGaugh (2000) has shown that the low amplitude of the second peak in the Boomerang and WMAP data is consistent with a no-CDM universe. Spergal comments on discrepancies in the WMAP data on both the largest and smallest scales, and Copi et al (2006) report on unexplained alignments in the data. The same authors conclude their 2007 report on the year 1-3 data "At the moment it is difficult to construct a single coherent narrative of the low microwave background observations. What is clear is that, despite the work that remains to be done understanding the origin of the observed statistically anisotropic microwave fluctuations, there are problems looming at large angles for standard inflationary cosmology". It is not presently possible to say whether these problems will be resolved by the teleconnection.

### 4  Friedmann Cosmologies with Local Gravity

#### *4.1 Static Coordinates*

Let two observers, Alf and Beth, be stationary with respect to each other in coordinates determined from light-speed (e.g. by the radar method). Let Beth be at a coordinate distance $r$ from Alf as measured in Alf's coordinates. Alf and Beth each measure the coordinate length of a short rod located at Beth's origin, and aligned on an axis with Alf and Beth. Using radar, Alf determines that the rod has a coordinate length $d$, while Beth determines a length $d'$ in her coordinates (figure 4). Because the rod is at Beth's origin, $d'$ is the proper length, of the rod.

When Alf and Beth try to align their spacetime diagrams (figure 5), maintaining synchronisation and constancy of the speed of light, they find a mismatch, because their clocks do not run at the same rate. As drawn, Beth's clock runs faster than Alf's by a factor $k > 1$ (the argument also holds for $k < 1$). Synchronisation requires that, for $k > 1$, Beth's coordinates are enlarged by the factor $k$, so that if Beth's coordinates are superimposed on Alf's, and aligned at Beth, Alf appears at a coordinate distance $r' = kr$, displaced from his position in his own coordinates.

The diagrams for measurement of the coordinate length of the rod are superimposed (figure 6). The proper time interval $2d'$ in Beth's coordinates corresponds to a time $2kd'$ in Alf's coordinates. The proper length $d'$ in Beth's coordinates appears with coordinate distance $d = d'/k$ in Alf's coordinates.

Beth turns her rod perpendicular to the axis from Alf (figure 7). Beth calculates that the angle subtended in tangent space by the rod at Alf is $\theta' = d'/r'$.

Imagine $n$ observers, $Beth_1$, $Beth_2$, …, $Beth_n$, with rods of proper lengths $d'_1$, $d'_2$, …, $d'_n$, positioned end to end, such that they form an unbroken circle of radius $r$ in Alf's coordinates (figure 8). The angles, $\theta_1$, $\theta_2$, ... $\theta_n$, subtended by the rods at Alf are not all equal. Let the redshift for $Beth_i$ be $k_i$. At each position $i$, $Beth_i$ applies a radial stretch with factor $1/k_i$. The coordinate systems are then superposed with Alf at the centre of the circle in each observer's coordinates. It is seen that $\theta_i = k_i \theta'_i$. It is already established that $r'_i = k_i r$. Then Alf's coordinate length for $Beth_i$'s rod is

$$d_i = r\theta_i = \frac{r'_i}{k_i} k_i \theta'_i = r'_i \theta'_i = d'_i,$$

equal to its proper length, as measured locally by $Beth_i$ (this is the defining condition for Schwarzschild coordinates).

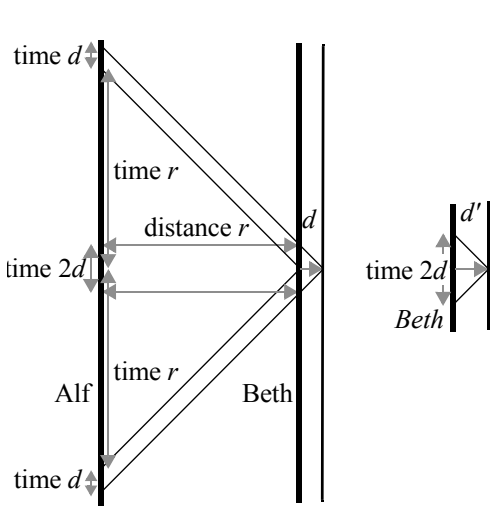

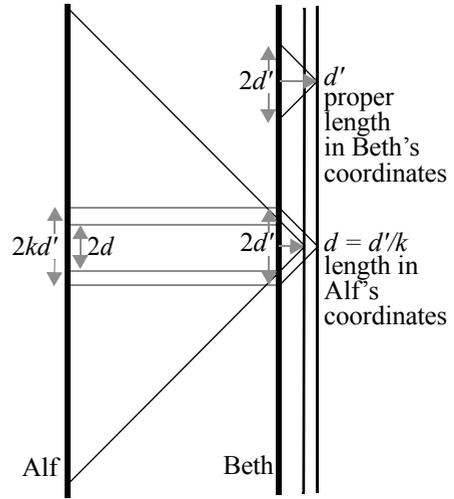

**Figure 4:** a) Length in Alf's coordinates. b) Length in Beth's coordinates.

**Figure 6:** (left) Measurement of the length of Beth's rod by Alf and Beth

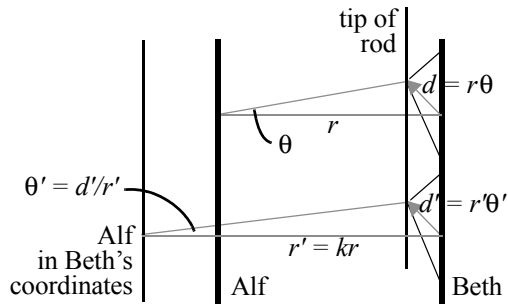

**Figure 7:** Length of Beth's rod perpendicular to a line from Alf to Beth

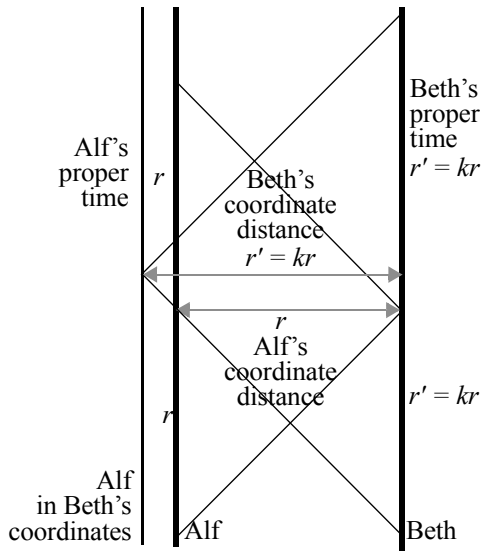

**Figure 5:** Synchronised coordinates determined by Alf & Beth using the radar method

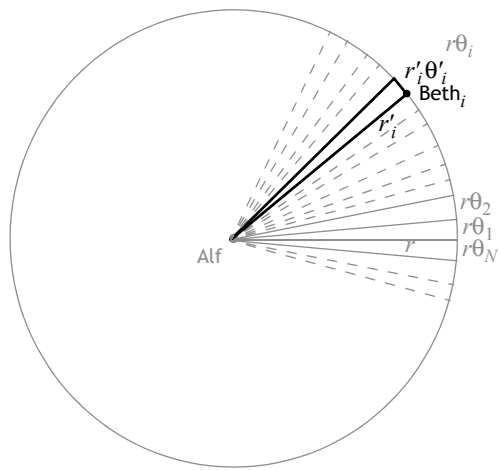

**Figure 8:** Space coordinates for *n* observers, Beth$_i$, with rods forming an unbroken circle about Alf, after applying a radial stretch $1/k_i$ to each Beth$_i$'s coordinates.

Thus, local to Beth, coordinate time $2kd'$ in Alf's coordinates corresponds to a proper time $2d'$, coordinate distance $d'/k$ corresponds to a proper distance $d'$, and angular distances are unchanged. The metric, $g$, takes a simple form. In spherical coordinates,

$$g_{ij} = \begin{bmatrix} 1/k^2 & 0 & 0 & 0 \\ 0 & -k^2 & 0 & 0 \\ 0 & 0 & -r^2 & 0 \\ 0 & 0 & 0 & -r^2\sin^2\theta \end{bmatrix}. \quad (4.1.1)$$

For stationary observers, if the redshift from A to B is $k_{AB}$, and the redshift from B to C is $k_{BC}$, then the redshift from A to C is $k_{AC} = k_{AB}k_{BC}$. Thus, for mutually stationary observers, the metric is a group.

### *4.2 Friedmann Cosmologies with Local Gravity*

In general relativity, time is determined from a clock locally. Although cosmic time is defined globally, its definition, based on Weyl's postulate, depends on the local time of many galaxies on geodesics from the Big Bang. The Friedmann models are based on the particular assumptions that the distribution of matter is homogeneous and isotropic. We expect these models to give a reasonable description of the universe at large scales, but they can only be approximate because they take no account of local mass distributions or of peculiar motions of galaxies and the orbits of stars within a galaxy. Friedmann models say nothing about what happens at smaller scales, at which matter is clearly not homogeneous. However, it is natural to think that local fluctuations in geometry due to the inhomogeneous local matter distribution can be treated as perturbations to a Friedmann model (at least where the gravitational field is not large).

Since it does not make sense to talk of expansion locally, $a(\tau)$ is a global parameter. We may define space-like hypersurfaces with $a(\tau) = \text{const}$ and define $\tau$ to be a global time parameter with $\tau = \text{const}$ on any surface with scale factor $a(\tau) = \text{const}$. Using Penrose coordinates, such that the coordinate speed of light is unity in the radial direction, the calculation of the general form of the metric for observers at constant $\rho$ goes through as for stationary observers, but now we have an additional factor of the expansion parameter. The line element has the form

$$ds^2 = a^2(\tau)(k^{-2}d\tau^2 - k^2 d\rho^2 - f^2(\rho)(d\theta^2 + \sin^2\theta d\phi^2)),$$

where the factor $k$ describes gravitational redshift. This description in trivially valid in a Friedmann model, and is clearly valid within regions with moderate gravity, but becomes singular at the event horizon of a singularity. There is no description of the region inside an event horizon in $\tau$-$\rho$ coordinates.

In contrast to a general description of the space-time manifold, $\tau$-$\rho$ coordinates with non-physical metric $\bar{h}$ have the important property that they are not affected by the matter distribution. A change in the matter distribution will affect the parameter $k$, and the wavelength of light, but not the property that light has constant unit speed in radial directions in these coordinates.

With the substitutions $dt' = a(\tau)d\tau$, $r' = a(\tau)\rho$, $\theta' = \theta$, and $\phi' = \phi$,

$$ds^2 = k^{-2}dt'^2 - k^2 dr'^2 - a^2(t')f^2\left(\frac{r'}{a(t')}\right)(d\theta'^2 + \sin^2\theta' \, d\phi'^2).$$

For $r' \ll a$,

$$ds^2 \approx k^{-2}dt'^2 - k^2 dr'^2 - r'^2(d\theta'^2 + \sin^2\theta' d\phi'^2),$$

in agreement with the metric for static coordinates, section 4.1. Locally Minkowski coordinates at the point $x$ with $r = 0$ are found by substituting $dt = k^{-1}dt'$ and $dr = kdr'$, $rd\alpha = r'd\theta'$, $r\sin\alpha \, d\beta = r'\sin\alpha \, d\phi'$. For small $r$,

$$ds^2 \approx dt^2 - dr^2 - r^2(d\alpha^2 + \sin^2\alpha \, d\beta^2).$$

Barred vectors are defined in $\tau$-$\rho$ coordinates as for a Friedmann cosmology. For a vector $x = (x^\tau, x^\rho, x^\theta, x^\phi)$, at $(\tau_1, A)$, the corresponding barred vector is

$$\bar{x} = \left(\frac{x^\tau}{2}, \frac{x^\rho}{2}, 2x^\theta, 2x^\phi\right).$$

For the vector $x = (x^t, x^r, x^\alpha, x^\beta)$ in locally Minkowski $t$-$r$ coordinates at $(t_1, A)$,

$$\bar{x} = \left(\frac{k}{2a}x^t, \frac{1}{2ak}x^r, \frac{2x^r}{x^{r'}}x^\alpha, \frac{2x^r \sin x^\alpha}{x^{r'}\sin x^\theta}x^\beta\right).$$

Alf formulates quantum states locally in Hilbert space at time $t_1$, and defines plane wave states at $t_1$ using

$$\langle x|p\rangle = \left(\frac{1}{2\pi}\right)^{3/2} e^{ix \cdot p}.$$

Quantum theory is reformulated globally using barred quantities, under the requirement that the inner product is preserved.

$$\langle \bar{x}|\bar{p}\rangle = \left(\frac{1}{2\pi}\right)^{3/2} e^{-i\bar{x}\cdot\bar{p}} = \left(\frac{1}{2\pi}\right)^{3/2} e^{-ix\cdot p} = \langle x|p\rangle.$$

This requires that $\bar{x} \cdot \bar{p} = x \cdot p$, and we define *barred momentum* by

$$\bar{p} = \left(\frac{ap^t}{2k}, \frac{akp^r}{2}, \frac{2r'p^\alpha}{r}, \frac{2r'\sin\theta p^\beta}{r\sin\alpha}\right).$$

We will be interested in weak fields and the transmission of photons over astronomical distances, for which momentum may be taken as radial up to the bending of light by a lens. It will be sufficient to use

$$\bar{p} = \left(\frac{ap^t}{2k}, \frac{akp^r}{2}, 0, 0\right).$$

### 4.3 Gravitational Redshift

Let A and B be sufficiently close that bending of light may be ignored. In coordinates with an origin at A, let the momentum of a photon passing from A at $t_1$ to B at $t_2$ be $p = (p^t, p^r)$. Let the redshift factor at A at time $t_1$ be $k_A$, and let the redshift factor at B at $t_2$ be $k_B$, and let $a_1 = a(t_1)$ and $a_2 = a(t_2)$. Then, barred momentum is

$$\bar{p} = \left(\frac{a_1 p^t}{2k_A}, \frac{a_1 k_A p^r}{2}, 0, 0\right).$$

Barred momentum is translated to $(\tau_2, B)$ in $\tau$-$\rho$ coordinates. Beth converts to locally Minkowski $t'$-$r'$ coordinates, and finds, with the removal of a factor of the expansion (section 2.3 and section 2.4),

$$(p^{t'}, p^{r'}, 0, 0) = \frac{a_2}{a_1}\left(\frac{a_1 k_B p^t}{a_2 k_A}, \frac{a_1 k_A p^r}{a_2 k_B}, 0, 0\right) = \left(\frac{k_B p^t}{k_A}, \frac{k_A p^r}{k_B}, 0, 0\right).$$

This is the same as is found by parallel displacement of momentum through a small distance in tangent space when the metric is given by (4.1.1)

$$ds^2 = k^{-2}d\tau^2 - k^2 d\rho^2 - f^2(\rho)(d\theta^2 + \sin^2\theta d\phi^2). \tag{4.3.1}$$

Thus, in locally Minkowski coordinates at A, gravitational redshift is given by a factor $k = (g_{00})^{-1/2}$, in agreement with standard general relativity.

### 4.4 Parallel Transport

The classical correspondence is found when there is sufficient information that states may be treated as being continuously measured. The quantum state is defined on a synchronous slice using quantum time, $t$. In order to describe a continuous classical motion, we must take a collection of synchronous slices, and treat each slice as the final state of one quantum step and as the initial state of the next quantum step, then allow the step size to go to zero, to obtain a foliation. At each stage of the motion, quantum theory is formulated in a space with a non-physical metric $\bar{h}$. Classical motion is determinate and may be described as an ordered sequence, $|f_i\rangle = |f(t_i)\rangle$, of effectively measured states at instances $t_i$ such that $0 < t_{i+1} - t_i < \delta$ where $\delta$ is sufficiently small that there is negligible alteration in predictions in the limit $\delta \to 0$. Each state, $|f_i\rangle$, is a multiparticle state in Fock space, which has been relabelled using mean properties determined by an effective classical measurement (by an effective measurement we mean that information is available such that the result of a measurement is determinate, whether or not a measurement is actually performed). For the motion between times $t_i$ and $t_{i+1}$, $|f_i\rangle$ may be regarded as the initial state and $|f_{i+1}\rangle$ may be regarded as the final state. Since the time evolution of mean properties is determinate, there is no collapse and $|f_{i+1}\rangle$ is the initial state for the motion to $t_{i+2}$. Classical formulae are recovered by considering a sequence of initial and final states in the limit as the maximum time interval between them goes to zero. In each stage of the motion momentum is parallel displaced using the teleconnection. This gives the same result as parallel displacement in tangent space (section 4.3). The cumulative effect of such infinitesimal parallel displacements is parallel transport. So the teleconnection between initial and final states in quantum theory leads to parallel transport in the classical domain (appendix A shows that this defines the Levi-Civita connection). The same argument holds for a classical beam of light, in which each photon wave function is localised within the beam at any time, and for a classical field which has a measurable value at each point where it is defined (up to the possible addition of an arbitrary gauge function).

### *4.5 Bending of Light*

The teleconnection predicts geodesic motion for a classical ray of light for which classical time and space coordinates can be determined at each point within a local reference frame. Thus, there is no change in the prediction of bending of light around the Sun. For bending by a distant gravitational lens, quantum wave effects are transmitted using parallel displacement of momentum in τ-ρ coordinates using the non-physical metric, $\bar{h}(x)$, given by (2.2.1):

$$d\bar{\sigma}^2 = 4(d\tau^2 - d\rho^2) - \frac{1}{4}f^2(\rho)(d\theta^2 + \sin^2\theta d\phi^2).$$

So, in the calculation of the deflection by a lens, we must halve the radial distance and double the angular distance, increasing the angle of deflection by a factor of four. Thus the mass required for a given strength of gravitational lens is a quarter of that which would be required in a standard no-CDM theory.

### *4.6 Quantum Coordinates*

If we wish to formulate quantum theory using coordinates with an origin which is not on a line of constant ρ, then we must retain the condition that the speed of light is unity.
**Definition:** *Quantum coordinates are defined such that the speed of light is unity in each direction from the origin, and the time coordinate is τ, and the non-physical metric is given locally by*

$$d\bar{\sigma}^2 = 4(d\tau^2 - d\rho^2) - \frac{1}{4}(d\theta^2 + \sin^2\theta d\phi^2). \tag{4.6.1}$$

τ-ρ coordinates are a special case of quantum coordinates, but it is not immediately clear how to generalise the angular coefficients in (2.2.1) to give a global description of the non-physical metric in the general case. Quantum coordinates have the property that they are not affected by the mass distribution. For small distances, parallel displacement of momentum in quantum coordinates is identical to parallel displacement in tangent space (locally Minkowski coordinates).

### *4.7 Circular Orbits*

Consider an inertial observer, Piers, at a point P at constant distance, $r$, from some arbitrary point, O, on a line of constant ρ. $r$ is sufficiently large that Piers' clock cannot be calibrated to a clock at O. Piers wishes to make P the centre of quantum coordinates, retaining the constraints that the speed of light is unity and time is τ. The Pioneer blueshift is equivalent to an acceleration of P toward O,

$$a_P = -H_0 c/2,$$

where there is a factor of two due to the non-physical metric. To take account of this acceleration, Piers introduces rotating coordinates with orbital velocity, $v_P$ (P is not at rest with respect to Piers). Because of the factors of $2^2$ in the non-physical metric (4.6.1), the orbital velocity corresponding to $a_P$ is subject to a factor of ¼, where one factor of ½ is from the time coordinate in the denominator of velocity, and another is from the transverse space coordinate. Thus, Piers defines coordinates in which P has an orbital velocity $v_P$ about O, such that

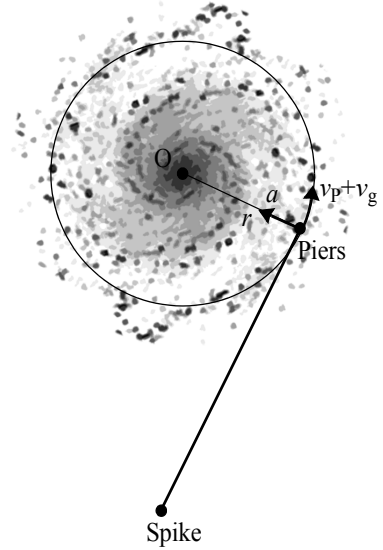

$$a_p = -\frac{16 v_P^2}{r} = -\frac{H_0 c}{2}.$$

Hence,

$$v_P = \pm\sqrt{\frac{H_0 c r}{32}}.$$

The matter distribution does not alter τ-ρ coordinates or $\bar{h}$. So, it does not alter the rotational velocity, $v_P$, required to define quantum coordinates. If the true orbital velocity of Piers in the gravitational field of a massive body at O, is $v_g$, the orbital velocity in quantum coordinates is

**Figure 9:**

$$v = v_g + v_P = \pm\sqrt{\frac{GM}{r}} \pm \sqrt{\frac{H_0 c r}{32}},$$

where the signs of each term are the same. The net acceleration toward O of a point at rest in quantum coordinates is

$$\frac{v^2}{r} = \frac{GM}{r^2} + \frac{\sqrt{GMH_0c/8}}{r} + \frac{H_0 c}{32}.$$

We use an interpretation in which the wave function describes a state of knowledge used for the calculation of probabilities. When a distant observer, Spike, observes photons from Piers (figure 9), he applies the classical constraint that there is no expansion in gravitationally bound systems. This removes the apparent acceleration $H_0 c/32$, leaving a spectral shift equivalent to total inward acceleration

$$\frac{v^2}{r} = \frac{GM}{r^2} + \frac{\sqrt{GMH_0c/8}}{r}. \tag{4.7.1}$$

The first term on the right hand side is simply the acceleration due to gravity and corresponds to Piers' orbital velocity about O. The second yields an illusory velocity corresponding to an illusory inverse acceleration law. When this shift is interpreted as an increase in orbital velocity it accounts for flattening of galaxies' rotation curves consistent with MOND, the phenomenological law proposed by Milgrom (1994) which retains Newton's square law for accelerations $\ddot{r} \gg a_M$ for some constant $a_M$ but replaces it with

$$\ddot{r} = -(GMa_M)^{1/2}/r \text{ for } \ddot{r} \ll a_M,$$

MOND gives a good match with data (see Sanders & McGaugh, 2002 for a review). (4.7.1) gives $a_M = H_0 c/8$. The empirical value of $a_M$ from observations on over a thousand stars is $1 \times 10^{-10}\,\text{ms}^{-2}$, giving $H_0 = 82\,\text{ms}^{-2}$. Using the teleconnection, a MONDian term arises from the treatment of redshift, not a change to Newtonian dynamics nor as evidence of cold dark matter haloes.

According to (4.7.1) there is no MONDian interpolation function between the regimes $\ddot{r} \ll a_M$, in which the Newtonian inverse square law dominates, and $\ddot{r} \gg a_M$, where the MONDian inverse law has been found. As a result, the predicted equivalent accelerations are larger than those predicted by MOND in the region $\ddot{r} \sim a_M$. This prediction agrees with studies on globular clusters by Scarpa et al. (2006), in which the flattening was observed at values of $\ddot{r}$ up to $2.1 \pm 0.5 \times 10^{-8}\,\mathrm{cm\,s^{-2}}$ and qualitatively agrees with the findings of de Blok & McGaugh (1998), who adjusted inclination so as to increase accelerations in the region of the interpolation function in order to obtain fits with MOND for a number of low surface brightness galaxies. These fits have not been recalculated.

In accordance with section 2.4, an illusory velocity is found for stars in bound orbits in the Milky Way, but not for planets in the solar system. This dichotomy is reflected in MOND, as well as in the Pioneer blueshift. Francis and Anderson (2009) considered whether flat rotation curves could be an artifact of an unmodelled component in spectral shift. In the absence of astrometric determinations of radial velocity, they applied a statistical test on a population of 20 574 Hipparcos stars inside 300 pc with known radial velocities and with accurate parallaxes in the New Hipparcos Reduction. The test rejected the null hypothesis, *there is no systematic error in spectrographic determinations of heliocentric radial velocity*, with 99.9993% confidence. In a separate test on metal-poor stars, they found tension between calculations of the orbital velocity of the Sun and three populations of halo stars inside and outside of a cone of 60° semi-angle from the direction of rotation. Tension cannot be removed with only systematic distance adjustments. They concluded that there is an unmodelled element in spectrographic determinations of heliocentric radial velocity with a probable cosmological origin, and proposed that this unmodelled component, rather than CDM or MOND, is responsible for the apparent flatness of galaxy rotation curves.

## 5 Conclusion

A formulation of quantum theory with remote initial and final states has been given for Friedmann cosmology with local fluctuations in gravity. Preliminary testing has been carried out on the empirical predictions this formulation, when light from distant bodies is treated in the quantum domain. The model differs from standard general relativity only in calculations involving cosmological redshift and lensing by distant objects. The tests described here could have falsified the formulation, but in fact support it

The formulation yields a closed expanding and contracting universe with approximately twice critical mass and with no cosmological constant, with an almost identical fit to supernova redshifts as the standard concordance cosmology. A modified age-redshift relation removes difficulties in models of galaxy evolution and predicts that new telescopes such as Herschel will find many large high-redshift galaxies.

Explanations have been given for the anomalous Pioneer blueshift and for galaxy rotation curves without resorting to cold dark matter or modifying Newtonian gravity. Distant lenses have a quarter of the mass required in standard general relativity. Missing mass can be accounted by a massive neutrino. CDM is not required. Direct evidence of an unmodelled component in radial velocity of local stars was found by Francis & Anderson (2009), who applied a straightforward statistical test of the prediction on a population of 20 574 local stars with accurate parallaxes and known radial velocities. The test rejected the null hypothesis, *there is no systematic error in spectrographic determinations of heliocentric radial velocity*, with 99.9993% confidence.

| Model | Standard | Teleconnection |
|---|---|---|
| Expansion rate | $\dot a/a = H$ | $\dot a/a = H/2$ |
| Critical density ($\Omega = 1$) | $3H_0^2/8\pi G$ | $3H_0^2/32\pi G$ |
| Topology | Open | Closed |
| $(\Omega,\Omega_k,\Omega_\Lambda)$ | ~(0.29,0,0.71) | ~(1.9,–0.9,0) |
| $\Omega_{stars}$ | 0.005~0.01 | 0.02~0.04 |
| $\Omega_B$ | 0.025~0.05 | 0.1~0.2 |
| Current age | ~13.7 Gyr | ~15.7 Gyr |
| Age at $z = 6$ | 0.9 Gyr | 3.2 Gyr |
| CDM haloes | inconsistent | absent |
| Pioneer blueshift | unexplained | $a_P = H_0 c$ |
| Galaxy rotation curves | CDM/MOND | Newtonian |
| Wave motion | curved space? | flat space |
| Classical motion | geodesic | geodesic |

Table 1: : Properties compared using ages from Ned Wright's calculator [http://www.astro.ucla.edu/~wright/ACC.html] assuming Hubble's constant h = 0.71.

The properties of a Friedmann cosmology with the teleconnection are summarized in table 1, based on Hubble's constant, $H_0 = 71$ km s$^{-1}$ Mpc$^{-1}$. The analysis of Francis & Anderson (2009) suggests that RR Lyrae distances are a little high, and hence that $H_0$ is a little low. $H_0 = 80$ km s$^{-1}$ Mpc$^{-1}$ would give better agreement with the standard age of the universe, and hence with the timescale for big bang nucleosynthesis, and better agreement with the observed values for the Pioneer drift, and MOND accelerations, equivalent to $H_0 = 90 \pm 14$ km s$^{-1}$ Mpc$^{-1}$ and $H_0 = 82$ km s$^{-1}$ Mpc$^{-1}$ respectively.

### Appendix A : Calculation of the Levi-Civita Connection

In many treatments the Christoffel symbols are said to define the connection and then used to define parallel transport. In this treatment, parallel displacement in tangent space is defined in the quantum domain, and reduces to parallel transport in the limit of small parallel displacement. This appendix is included for completeness and shows that the result is the same.

#### A1 Christoffel Symbols

Consider a region in which the metric field is $g_{ab}$.

**Definition:** *Christoffel symbols of the first kind:*

$$\Gamma_{abc} = (g_{ab,c} + g_{ac,b} - g_{cb,a})/2 .$$

**Definition:** *Christoffel symbols of the second kind:*

$$\Gamma^a_{bc} = g^{ad}\Gamma_{dbc} .$$

Clearly Christoffel symbols are symmetric in the last two suffixes, and satisfy

$$\Gamma_{abc} + \Gamma_{bac} = g_{ab,c} .$$

The purpose of Christoffel symbols is to enable us to calculate the effect of parallel displacement without reference to a non-physical tangent space. Tangent space at $x$ is defined with non-physical metric $h$ using primed coordinates.

$$g_{ab}(x) = k_a^{m'}(x)k_b^{n'}(x)h_{m'n'} .$$

$$g_{ab}(x+dx) = k_a^{m'}(x+dx)k_b^{n'}(x+dx)h_{m'n'} + O(\max(dx^i)) .$$

$$g_{ab,c}(x) = k_{a,c}^{m'}(x)k_b^{n'}(x)h_{m'n'} + k_a^{m'}(x)k_{b,c}^{n'}(x)h_{m'n'} . \quad (A1.1)$$

From Clairaut's Theorem, the partial derivative of the transformation matrix is symmetrical in its lower indices, $k_{c,b}^{m'} = k_{b,c}^{m'}$. Interchange $a$ & $c$ and $b$ & $c$ in (A1.1).

$$g_{cb,a} = k_{c,a}^{m'}k_b^{n'}h_{m'n'} + k_c^{m'}k_{b,a}^{n'}h_{m'n'} ,$$

$$g_{ac,b} = k_{a,b}^{m'}k_c^{n'}h_{m'n'} + k_a^{m'}k_{c,b}^{n'}h_{m'n'} .$$

Then,

$$\Gamma_{abc} = (g_{ab,c} + g_{ac,b} - g_{cb,a})/2 = k_a^{m'}k_{b,c}^{n'}h_{m'n'} .$$

#### A2 Parallel Displacement

Parallel displacement of a vector $p$ from $x$ to $x+dx$ in tangent space keeps the primed components constant, $p^{m'}(x+dx) = p^{m'}(x)$. Multiply by $k_b^{n'}(x+dx)h_{m'n'}$, to lower the index and convert to unprimed coordinates.

$$p_b(x+dx) = p^{m'}(x)k_b^{n'}(x+dx)h_{m'n'} + O(\max(dx^i)) .$$

$$p_b(x+dx) = p_b(x) + p^{m'}(x)k_{b,c}^{n'}(x)dx^c h_{m'n'} + O(\max(dx^i)) .$$

Then, ignoring terms $O(\max(dx^i))$,

$$dp_b = p^{m'}(x)k_{b,c}^{n'}(x)dx^c h_{m'n'} = p^a(x)k_a^{m'}(x)k_{b,c}^{n'}(x)dx^c h_{m'n'} .$$

Substituting the Christoffel symbol eliminates the dependency on tangent space.

$$dp_b = \Gamma_{abc}p^a dx^c .$$

Raise and lower indices to find the standard formula for infinitesimal parallel displacement referring to covariant components.

$$dp_b = \Gamma^a_{bc} p_a dx^c.$$

For a second vector $q$, $p \cdot q$ is invariant.

$$d(p_b q^b) = 0.$$

$$dp_b q^b + p_b dq^b = 0.$$

$$p_a \Gamma^a_{bc} dx^c q^b + p_a dq^a = 0.$$

Since this is true for all $p_a$, we find the standard formula for infinitesimal parallel displacement referring to contravariant components,

$$dq^a = -\Gamma^a_{bc} q^b dx^c$$

### A3 The Covariant Derivative

We may parallel displace a vector from $x$ to a nearby point, $x + dx$ and it becomes a vector defined at $x + dx$. We can then define the covariant derivative using vector addition, and the result will be a tensor. For contravariant and covariant vector fields, $p_b$ and $q^a$, the covariant derivative is defined by

$$p_{b\,;\,i} : x \to p_{b\,;\,i}(x) = \lim_{dx^i \to 0} \frac{p_b(x + dx^i) - (p_b(x) + \Gamma^a_{bc} p\, dx^i)}{|dx^i|},$$

$$q^a_{\,;\,i} : x \to q^a_{\,;\,i}(x) = \lim_{dx^i \to 0} \frac{q^a(x + dx^i) - (q^a(x) - \Gamma^a_{bc} q^b dx^i)}{|dx^i|},$$

where $dx^i$ is a small vector along the $i$-axis. One may see that $i$ is a vector index from first principles, or by using the result that the partial derivative of a scalar is a vector for each value of $a$ and $b$. Hence we find the standard forms,

$$p_{b\,;\,i} = p_{b,i} - \Gamma^a_{bi} p_a, \text{ and } q^a_{\,;\,i} = q^a_{,i} + \Gamma^a_{bi} q^b.$$